# Automation of Smartphone Traffic Generation in a Virtualized Environment

Tanya Jha
Rashmi Shetty

## Abstract

Scalable and comprehensive analysis of rapidly evolving mobile device application traffic is extremely important but a challenging problem for the Deep Packet Inspection (DPI) engines to perform effective policy management. We present a test framework in which a test driver can automate/orchestrate traffic generation by invoking appropriate method (intent) of real mobile applications (as opposed to traffic replay) in regression or functional testing of mobile application traffic analysis engines in a virtualized environment, without real hardware. We demonstrate the concept by automating a real-time Skype call through a DPI engine in a virtual test setup using Android VMs.

## I. Introduction

The last decade has seen the transformation of mobile phones from being mere communication devices to multipurpose gadgets that has resulted in their immense popularity. The landscape of mobile devices like smartphones and tablets has witnessed a tectonic shift with its usage rising by the hour. TechCrunch estimates that by the year 2020, there would be a massive 6.1 billion smartphone users globally, overtaking basic fixed phone subscriptions. According to Search Engine watch, in early 2014, Internet usage on mobile devices exceeded PC usage.

This smart phone revolution could be attributed to the wide range of digital services offered by them. Presence of digital wallets, easy and safe mobile bank transactions, finding a cab, pizza or bride for wedding at the click of a button, myriad games like Temple Run, Asphalt, Angry Birds and Candy Crush are some of the many features of these smart devices that have made our lives easier and more entertaining.

Recent surveys reveal that smart phone applications generate about 82% of Mobile IP. Such applications are constantly evolving with every version. Traffic analysis (especially DPI) is a challenge without access to applications running on a real phone platform. Hence here we propose a test framework in which a test driver is used to automate the app traffic generation. This is done by invoking android intents from the test driver to simulate the traffic and analyze it in a virtualized environment. Inspection of the traffic, collection of statistics and application of policies would be simplified.

This paper is organized as follows: Section 2 describes the existing solutions that are currently in use. Section 3 highlights on the challenges with the existing solutions and the need to move to a fully virtualized model. Section 4 elaborates on the proposed solution. Section 5 provides details of the implementation. Section 6 concludes the paper.

## II. Existing Solutions

1) Existing solutions replay packet captures using physical hardware. The problem with this is that the different versions of the applications released cannot be tested immediately.
2) Due to the replaying of packet captures the behaviour of the application cannot be controlled.
3) Another problem is the cost of hardware required for this purpose.

## III. Challenges:

- The explosion of smartphones, tablet computers, and other wireless-enabled devices, coupled with the availability of thousands of new applications that leverage IP-based mobile data networks, is creating a new and challenging environment for the wireless service provider. This environment is a lot more transient and unpredictable than traditional mobile voice networks and presents complex challenges for service providers to inspect/enforce policies on the traffic.
- Today, service providers find it difficult to identify and characterize the impact that specific sources (i.e., devices, local and Internet applications, etc.) have on network capacity, performance, and security. Traditional radio management tools can indicate when performance is bad or when a certain capacity is being exceeded, but they do not explain why or which applications or devices are causing the problem. Besides, these operations involve large expenses.
- "Deep Packet Inspection" is being implemented to provide a range of services from content filtering, to web application firewalls, application aware policy enforcement, and others. One of the challenges faced by a DPI engine is the practical limitations to keep up with the constantly evolving mobile application traffic patterns. Secondly such applications require special platform (for example a phone) to run.
- Critical applications need bandwidth prioritization while social media and gaming applications need to be bandwidth throttled or completely blocked. Stateful packet inspection firewalls used in many organizations rely on port and protocol; they cannot solve the problem because they are not able to identify applications

## IV. Proposed Solution

We explore the possibilities of implementing a Driver Interface that can automate the Android Virtual Machines acting as smartphones. This Linux Driver Interface, exploiting Android's intent functionality, would help in automating the testing of traffic generation by popular applications and also enable testing of DPI engine against different versions of such apps (E.g. Skype) in a virtualized environment using Android VM. The proposed solution requires no real hardware (except to host the VMs) and the latest app versions can be deployed immediately after their release (via app-store/ market update)

## V. Implementation

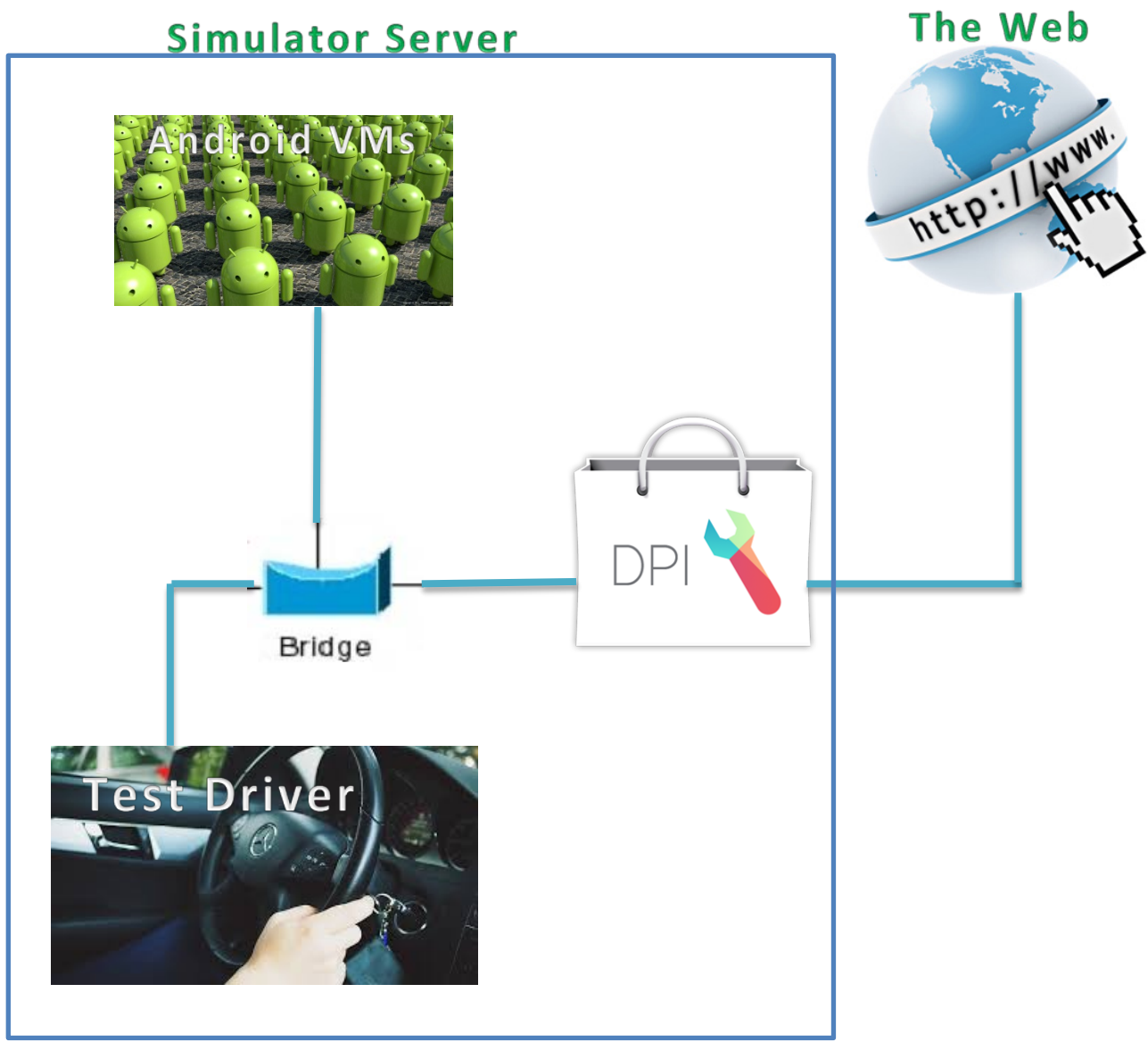

The idea is to create a topology as depicted in the above diagram. On a single host we can bring up Android Virtual Machines, Linux VM and a Router Simulator. Android VMs on the left representing smartphones are connected to the external network (cloud) via the router simulator. This router acts as the default IP gateway for the data sent and received from Android VMs. Routing would be enabled on both Android and Linux VM so that they can communicate with each other and when they do, we can perform DPI operations on the traffic.

The iso images for the Android VMs could be downloaded from android developer's site. The Linux VM iso images are available on the internet. This test driver, in addition to being used for the complete automation of android bring up, it also acts as a DHCP server for the Android VMs. Whenever a new android VM comes up, it does a dhcp request and is then granted an address from the Linux test-tool. After bringing up the 2 VMs, a virtual bridge is to be created to hook up Android VMs and the test driver to the DPI tools enabled router simulator. Android VMs would now be ready to stream in live data.

To automate the android VMs the scripts can be written on a Linux tool. Along with the automation of Android bring up, Linux VM also serves as a DHCP server. Since traffic pattern analysis is done on many devices, every time an Android VM runs an app, it is possible to track the device that has triggered the application using its IP address. Linux VM is made to run the DHCP server and Android VM gets an address from a range of IP addresses set aside in /etc/dhcpd.conf file.

Dhcpd.leases file contains the latest IP address that was assigned to a newly created Android VM with other information like lease duration etc. We assume that the lease will be given only for an Android VM and no other VM other than an Android will be making a DHCP request.

A script will be running continuously in the Linux VM and will be checking the dhcpd.leases file at regular intervals to see if a new Android VM has come up. Only the IP address is grepped from the lease file. This IP address can be used to connect to the android VM using ADB shell.

ADB (Android Debugging Bride) shell is a command line tool that allows communication with the Android VM. The adb shell has 'am commands' that allows the connection to an android device (in this case, the VM) and then use it to install applications and trigger the same. Once the Android SDK is downloaded and installed, its platform-tools packages are enabled which contain ADB and other utilities.

On the Linux VM we install adb. Using the adb shell we communicate with the Android VM. We go to platform-tools in the adt-bundle-linux-x86 folder and then execute the commands for the connected device. We can generate a list of attached emulators/devices using the devices command:

*adb devices*

If the device has not been attached, use
- For emulator :

  *adb –s emulator5556*
- For virtual machine :

  *./adb connect ip_address:5555*

  (Using the IP address assigned to the VM)

Different operations can be performed from this shell thereby allowing a complete automation of the android bring up: Applications can be installed by downloading their corresponding .apk files from the internet and installing on the VM. E.g.

*./adb install - rTwitter_3.0.1.apk*

Within an adb shell, we can issue commands with the activity manager (am) tool to perform various system actions, such as start an activity, force-stop a process, broadcast an intent, modify the device screen properties, and more. While in a shell, the syntax is:

*am <command>*

We can also issue an activity manager command directly from adb without entering a remote shell. For example:

*adb shell am start –a android.intent.action.VIEW*

Within an adb shell, we can issue commands with the package manager (pm) tool to perform actions and queries on application packages installed on the device. While in a shell, the syntax is:

*pm <command>*

We can also issue a package manager command directly from adb without entering a remote shell. To uninstall an application you can use the pm command. For example:

*adb shell pm uninstall com.example.MyApp*

An android application can be written to take advantage of the intent function of applications. So test applications can easily be written to initiate a skype call, or launch Angry birds towards pigs. These applications will greatly help automation especially in headless mode, where no human interaction is available to invoke. Such test applications can be run as a part of /system/etc/init.d scripts or as cron job or can be invoked remotely. In addition, automated creation of scores of such VMs can help with creating real world scalable traffic patterns.

As an example we wrote an application Appmanager. It is an android application that is used to invoke other applications. The applications used by us are Facebook, twitter and skype. Other applications can also be added to the Appmanager. Appmanager has a main class(MainActivity.class) and separate classes for the applications that we wish to use (e.g fb.class).

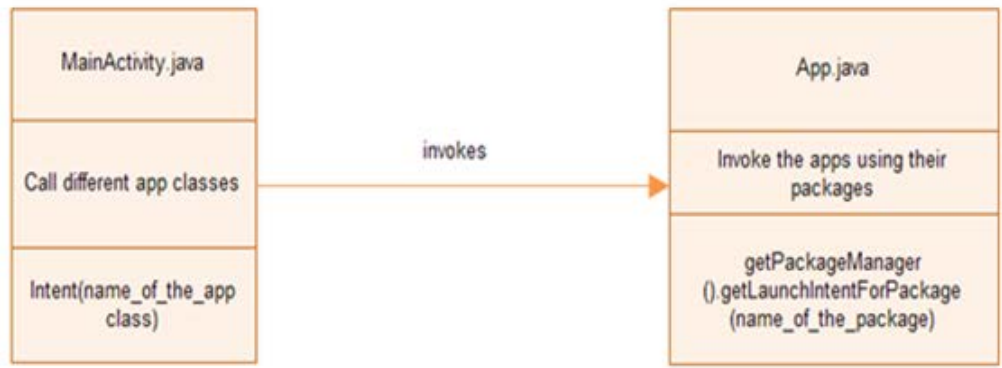

The Appmanager has a main class that has a list of classes for the different applications. These classes are called one after another. In this main class we create intent by passing parameters consisting of this mainactivity and the activity of the application that we wish to start. We then pass this intent to the startActivity() function.

In the individual classes for the separate applications, we create an intent for that particular application using the package name.

*Intent appIntent = getPackageManager().getLaunchIntentForPackage("Package Name");*

If we have the app installed on a device and can connect using adb, you can launch adb shell and execute *pm list packages -f*, which shows the package name for each installed apk. We then pass this intent to the startActivity() function that will then start the application. In the mainActivity class after starting an activity we introduce some delay so that we can view the application opening on the VM. Then the current activity is closed and then the new activity is called.

## VI. Conclusion

In this paper we look into the possibilities of running virtual machines with router simulator as IP gateway inspecting/enforcing policies on the traffic and demonstrate possibilities of integrating it as automated test-suite in a virtual environment. The android phones are simulated as virtual machines.

With this setup, it is possible to simulate 100s of phones as virtual machines inside a single server, and they can all be connected to a single routing device, simulating a whole network of phones and analysis can be done on aggregation of traffic.

This setup will help in passively monitoring, in real-time, every subscriber's data experience while automatically analyzing and identifying the root-cause issues such as anomalous events (e.g heavy users, signaling overloading, security threats, etc) that are contributing to a subscriber's degraded experience.